\documentclass[twocolumn,showpacs,preprintnumbers,%
amsmath,amssymb]{revtex4}

\usepackage{graphicx}
\usepackage[colorlinks]{hyperref}

\providecommand*{\bra}[1]{\langle#1\rvert}
\providecommand*{\ket}[1]{\lvert#1\rangle}
\providecommand*{\abs}[1]{\lvert#1\rvert}
\DeclareMathOperator{\tr}{tr}

\begin{document}

\title{%
  Mapping Kitaev's quantum double lattice models \\
  to Levin and Wen's string-net models}

\author{Oliver Buerschaper}

\author{Miguel Aguado}

\affiliation{%
  Max-Planck-Institut f\"ur Quantenoptik.  Hans-Kopfermann-Stra\ss{}e
  1, D-85748 Garching, Germany}

\date{\today}



\begin{abstract}
  We exhibit a mapping identifying Kitaev's quantum double lattice
  models explicitly as a subclass of Levin and Wen's string net models
  via a completion of the local Hilbert spaces with auxiliary degrees
  of freedom.  This identification allows to carry over to these
  string net models the representation-theoretic classification of the
  excitations in quantum double models, as well as define them in
  arbitrary lattices, and provides an illustration of the abstract
  notion of Morita equivalence.  The possibility of generalising the
  map to broader classes of string nets is considered.
\end{abstract}

\maketitle

\section{\label{sec:intro}Introduction}

Quantum lattice models governed by local Hamiltonians exhibit a wealth
of phases that sometimes escapes the local group analysis of
symmetries underlying Landau's paradigm for second order phase
transitions.  Topological phases \cite{WenOrder} are a remarkable
instance of such exotic behaviour, where the effective field theory
controlling the long-distance properties is a topological quantum
field theory (TQFT) \cite{tqft}.  Such phases arise in the fractional
quantum Hall effect (and also in $p$-wave superconductors), but they
also find explicit realisation in a number of lattice models,
intensely studied both for their theoretical properties and because of
the proposal by Kitaev \cite{Kitaev} that they would make
intrinsically fault-tolerant quantum memories and computers.

Among these lattice models, the class of quantum double (QD) models,
corresponding to discrete lattice gauge theories, was introduced as
candidates for quantum memories and computers in \cite{Kitaev}.  The
more comprehensive class of string-net (SN) models was defined in
\cite{LevinWen}.  These models have a characteristic form of
frustration-free Hamiltonian as a sum of mutually commuting
projectors.  It is these two classes of models that we will deal
with in this paper.  On the other hand, interesting models with more
intricate Hamiltonians have been proposed: let us mention the extended
Hubbard model \cite{extdHubbard}; Kitaev's honeycomb model
\cite{Kitaev:2005}, featuring two-body interactions; colour codes
\cite{colourcodes}, and generalised quantum doubles \cite{genkitaev}.

The study of topological order brings together many abstract
mathematical concepts, from fields ranging from quantum groups to
category theory, yet the simplest topological lattice models offer a
very direct physical bridge to these concepts.  For instance, the
representation theory of quasitriangular Hopf algebras governs the
classification of excitations in QD models (to be discussed in section
\ref{sec:qdoubles}), and fusion categories are the starting point for
SN models.

It is desirable to understand the interrelations among different
lattice constructions of systems with topological order.  As argued in
\cite{LevinWen}, two-dimensional SN models encompass all
\emph{doubled} topological phases, and it is understood that the
discrete gauge theory phases described by QD models should be
contained in the class of SN models.  In section \ref{sec:mappingqd},
we show how this happens and identify quantum doubles with a subclass
of string-net models (a construction, to the best of our knowledge,
not so far made explicit in the literature).  In more abstract terms,
this is an example of Morita equivalence\footnote{We are indebted to
  Zhenghan Wang for pointing this out.} (the origin of this concept
can be found in \cite{morita}; see, e.~g., \cite{andersonfuller}),
whereby the local degrees of freedom in the lattice may be seen as
objects in a category, and the physical excitations, equivalent in
both cases, correspond to a representation category.

The plan of the paper is as follows: We briefly introduce the quantum
double models in section \ref{sec:qdoubles}, and the string-net models
in section \ref{sec:stringnets}.  The mapping $\text{QD} \rightarrow
\text{SN}$ is discussed in section \ref{sec:mappingqd}, and in section
\ref{sec:conclusions} we present our conclusions, and a preview of
work in progress concerning generalisations of this construction.

\section{\label{sec:qdoubles}%
  Quantum double models}

Quantum double lattice models are a direct translation into the
lattice setting of gauge theories with a finite, discrete gauge group
$G$ \cite{Wegner, Bais}.  They were proposed by Kitaev as quantum
memories in \cite{Kitaev}, with the purpose of obtaining anyonic
excitations capable of universal quantum computation by braiding.
Mochon \cite{Mochon} proved that already the model with the smallest
non-Abelian gauge group $G = \mathrm{S}_3$ is universal in this sense,
assuming certain `magic states' can be prepared.  While exhibiting
rich non-Abelian anyonic excitations, these models share some of the
simplicity of their first instance, the toric code \cite{Kitaev} (in
particular, topological sectors have integral quantum dimensions).
For a recent review using group theoretic language, see
\cite{genkitaev}.

The quantum double model based on a finite group $G$, the $\mathrm{D}
(G)$ model for short, is defined on an \emph{arbitrary} planar lattice
$\Lambda$, with local degrees of freedom associated with oriented
edges $e$, the local Hilbert space $\mathcal{H}_e$ being
$\abs{G}$-dimensional, with $\abs{G}$ the order of $G$.  The
orthonormal computational basis is labeled by group elements,
$\mathcal{B}_{\mathbf{B}}=\{\ket{g}\vert g\in G\}$. The Hilbert space
$\mathcal{H}_{e^\ast}$ for the reversed edge ${e^\ast}$ is identified
with $\mathcal{H}_e$ via the isomorphism $\ket{g}_e \mapsto \ket{
  g^{-1} }_{e^\ast}$.

The Hamiltonian is of the form
\begin{equation}\label{qdouble:hamiltonian}
  H^{ \mathrm{QD} }
=
 -
  \sum_v A_v^{ \mathrm{QD} }
 -
  \sum_p B_p^{ \mathrm{QD} }
  \; ,
\end{equation}
where $p$ runs over the plaquettes of $\Lambda$, $v$ over the
vertices, and the $A_v^{ \mathrm{QD} }$ and $B_p^{ \mathrm{QD} }$ are
mutually commuting projectors acting on the edges surrounding each
vertex or plaquette. Their action on computational basis states is
defined by the following:
\begin{itemize}
\item %
  For each vertex $v$, orient its adjacent edges ($\deg(v)$ in number)
  such that they point towards $v$. Denote a ket in the computational
  basis for this set of oriented edges as $\ket{ \{ g_i \} } = \ket{
    g_1 } \otimes \ket{g_2} \otimes ...\otimes \ket{ g_{\deg v} }$,
  where $i = 1, \, \ldots, \, \deg(v)$ labels the edges (cf.~figure
  \ref{fig:vertexqd}).
  \begin{figure}
  \centering
  \includegraphics[scale=.8]{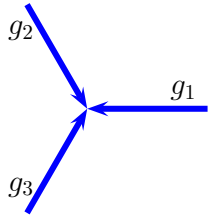}
  \caption{\label{fig:vertexqd}%
    Convention for the definition (\ref{qdouble:vprojector_orig}) of
    operators $A_v^{ \mathrm{QD} }$ in quantum double models.  A
    trivalent vertex is shown, but the generalisation to arbitrary
    vertices is straightforward.  The operator is symmetric in the
    different edges.}
  \end{figure}
  Then the vertex projector acts as
  \begin{equation}\label{qdouble:vprojector_orig}
    A_v^{ \mathrm{QD} }
    \ket{ \{ g_i \} }_v
  =
    \frac{ 1 }{ \abs{G} }
    \sum_{ h \in G }
    \ket{ \{ g_i h^{-1} \} }_v,
  \end{equation}
  i.e., it is a simultaneous right multiplication averaged over the
  group.
\item %
  For each plaquette $p$ pick an arbitrary adjacent vertex $v_0$ as
  starting point and denote $\ket{ \{ g_j \} }_p$ a ket in the
  computational basis with $j = 1, \, \ldots, \, s$ labelling the
  ordered edges of $p$ along the counterclockwise path starting and
  ending at $v_0$ (cf.~figure \ref{fig:plaquetteqd}).
  \begin{figure}
  \centering
  \includegraphics[scale=.8]{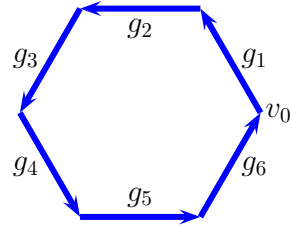}
  \caption{\label{fig:plaquetteqd}%
    Convention for the definition (\ref{qdouble:pprojector_orig}) of
    operators $B_p^{ \mathrm{QD} }$ in quantum double models.  A
    hexagonal plaquette is shown, but the generalisation to arbitrary
    plaquettes is straightforward.  The definition of $B_p^{
      \mathrm{QD} }$ does not depend on the choice of $v_0$ or
    orientation of the loop.}
  \end{figure}
  Then the plaquette projector acts as
  \begin{equation}\label{qdouble:pprojector_orig}
    B_p^{ \mathrm{QD} }
    \ket{ \{ g_j \} }_p
  =
    \delta (g_1 \cdots g_s, \, e )
    \ket{ \{ g_j \} }_p
    \; ,
  \end{equation}
  projecting onto states where the product $g_1 \cdots g_s$ equals the
  identity element $e \in G$ ($\delta$ is a Kronecker delta). Both the
  choice of the starting vertex $v_0$ and of the orientation are
  immaterial, but the order is crucial.
\end{itemize}

Breakdown of any of the ground level constraints $A_v^{ \mathrm{QD} }
= +1$ and $B_p^{ \mathrm{QD} } = +1$ takes the system to excited
levels, i.e., to the eigenspaces with eigenvalue $-1$ of the vertex
and plaquette operators.  The structure of the excitations can be
understood as the appearance of quasiparticles in the regions where
constraints are violated.  These quasiparticles possess mutual anyonic
statistics, which are non-Abelian if the group $G$ is non-Abelian;
they can be classified into superselection sectors (topological
charges) given by irreducible representations of the \emph{quantum
  double} $\mathrm{D} (G)$, a quasi-triangular Hopf algebra
constructed from the group algebra of the finite group $G$ by
Drinfel'd's quantum double construction \cite{qdoublealgebra} (a clear
introduction for physicists is \cite{lahtinen}):
\begin{itemize}
\item %
  Magnetic charges, or fluxes, correspond to the violation of a
  plaquette constraint, $B_p^{ \mathrm{QD} } = -1$.  They are given by
  conjugacy classes of elements of $G$.
\item %
  Electric charges correspond to the violation of a vertex constraint,
  $A_v^{ \mathrm{QD} } = -1$, they are given by irreducible
  representations (irreps) of $G$.
\item %
  Dyonic charges correspond to the violation of both a plaquette and a
  neighbouring vertex constraint, they are given by a conjugacy class
  $C$ of $G$ and an irrep of the centraliser of $C$ in $G$.
\end{itemize}
In addition, each topological sector has a number of internal states
(given, for instance, by representatives of a conjugacy class in the
case of magnetic charges, and by irrep basis vectors in the case of
electric charges.)  The topological charge can be measured for a
region by operations along its boundary, and can be defined more
generally for any closed loop, not necessarily the boundary of a
region.  Projectors onto the different superselection sectors can be
constructed with Kitaev's \emph{ribbon operators} \cite{Kitaev}.

\section{\label{sec:stringnets}%
  String-net models}

The class of string-net lattice models was introduced in
\cite{LevinWen}.  The intuition behind the construction is that
topological lattice Hamiltonians are infrared fixed points of some
renormalisation group procedure (this has since been made explicit in
\cite{AguadoVidal} and \cite{Koenig} using entanglement
renormalisation techniques).  String-net models in two dimensions
describe all \emph{doubled} topological phases, including interesting
cases where the excitations are given by the double semion model or
the double Fibonacci models.

The starting point is a honeycomb lattice (or any planar lattice with
\emph{trivalent} vertices) with local degrees of freedom along its
(oriented) edges.  For any oriented edge, an orthonormal basis $\{
\lvert a \rangle \}_e$ is labelled by the charges $1, \, a, \, b,
\ldots \in M$ in an anyon model (more precisely, a unitary fusion
category) featuring particle-antiparticle duality $a \leftrightarrow
a^\ast$, a set of fusion rules $a \times b \rightarrow \sum_c N_{ab}^c
c$, and fusion/splitting linear spaces such that the recoupling
isomorphisms for a $2 \rightarrow 2$ process are given by $F$-symbols
\cite{Kitaev:2005}.  The bases for the two different orientations $e,
e^\ast$ of the same edge are related by duality, $\lvert a
\rangle_{e^\ast} = \lvert a^\ast \rangle_e$.

\begin{figure}
\centering
\includegraphics[scale=.8]{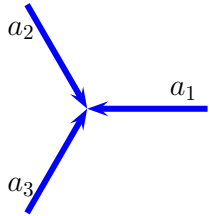}
\caption{\label{fig:vertexsn}%
  Convention for the definition (\ref{stringnets:vertexsn}) of
  operators $A_v^{ \mathrm{SN} }$ in string-net models.  The operator
  is symmetric in the different edges.}
\end{figure}

The string-net Hamiltonian $H^{\text{SN}}$ is constructed from the
data of the fusion category:
\begin{equation}\label{stringnets:hamiltonian}
  H^{\text{SN}}
=
 - \, \sum_v A^{\text{SN}}_v
 - \, \sum_p B^{\text{SN}}_p
  \;,
\end{equation}
where $v$ labels the vertices of the lattice, and $p$ its plaquettes.
Each vertex term $A^{\text{SN}}_v$ is a three-body projector (acting
on the three vertices incoming to $v$) favouring fusion rules.  That
is, they project out vertex configurations $\lvert a, \, b, \, c
\rangle_v = \lvert a \rangle_{e_1} \otimes \lvert b \rangle_{e_2}
\otimes \lvert c \rangle_{e_3}$ with $N_{ab}^{c^\ast} = 0$ (the
orientations of the $e_j$ are chosen pointing towards $v$, see figure
\ref{fig:vertexsn}).
We write
\begin{equation}\label{stringnets:vertexsn}
  A^{\text{SN}}_v
  \lvert a, \, b, \, c \rangle_v
  =
  \sum_{ a, b, c \in M }
  \delta( a \times b \times c \rightarrow 1 )
  \lvert a, \, b, \, c \rangle_v
  \; ,
\end{equation}
with the obvious meaning for the delta (if $a \times b \rightarrow
c^\ast + \ldots$, then $a \times b \times c \rightarrow 1 + \ldots$).

\begin{figure}
\centering
\includegraphics[scale=.8]{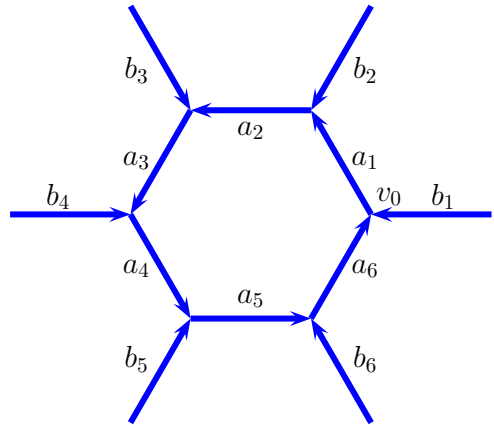}
\caption{\label{fig:plaquettesn}%
  Convention for the definition (\ref{stringnets:plaquettesnfull}) of
  operators $B_p^{ \mathrm{SN} }$ in string-net models.  The
  definition of $B_p^{ \mathrm{SN} }$ does not depend on the choice of
  $v_0$ or orientation of the loop.}
\end{figure}

Plaquette projectors $B^{\text{SN}}_p$ can be constructed from the
$F$-symbols of the fusion category.  Pictorially, they correspond to
the introduction of loops of different labels within the corresponding
plaquette and then subsuming the loop into the original lattice by
means of recoupling ($F$-)moves (this can be made precise in the
fattened lattice picture of the string-nets \cite{LevinWen};
technically, the loop is to be introduced enclosing the puncture for
plaquette $p$).  We denote the operation associated with a loop with
label $c$ as $B^{\text{SN}}_p ( c )$, and then
\begin{equation}\label{stringnets:plaquettesn}
  B^{\text{SN}}_p
=
  \sum_{ c \in M }
  \frac{ d_c }{ \mathcal{D}^2 }
  B^{\text{SN}}_p ( c )
  \; ,
\end{equation}
where $\{ d_a \}$ are the \emph{quantum dimensions} of the different
labels in the fusion category, and $\mathcal{D}^2 = \sum_{ b \in M }
d_b^2$ is the \emph{total quantum dimension}.  The explicit action of
the $B_p (a)$'s is spelt out in \cite{LevinWen}; the net result for
(\ref{stringnets:plaquettesn}) is
\begin{align}\label{stringnets:plaquettesnfull}
\nonumber
  B^{\text{SN}}_p
&
  \lvert \{ a_i \}; \, \{ b_j \}  \rangle_p
\\
&=
  \sum_{ c, \, \{ a' \} }
  \frac{ d_c }{ \mathcal{D}^2 }
  \Bigg(
    \prod_{ \ell=1 }^6
    F^{
       b_j {a'_\ell}^\ast a'_{\ell-1}
    }_{
       c    a_{\ell-1}    a_\ell^\ast
    }
  \Bigg)
  \lvert \{ a'_i \}; \, \{ b_j \}  \rangle_p
  \; ,
\end{align}
where $\{ a_i \} = \{ a_1, \, \ldots, \, a_6 \}$ label the edges along
a loop following the boundary of the plaquette (taken hexagonal for
definiteness) counterclockwise, and $\{ b_i \} = \{ b_1, \, \ldots, \,
b_6 \}$ is the configuration of the edges immediately neighbouring
these and pointing towards $p$ (cf.~figure \ref{fig:plaquettesn}).
The action of $B^{\text{SN}}_p$ depends on the $b$'s but only changes
the $a$ labels; all in all, a twelve-body operator.

All plaquette and vertex constraints can be seen to commute with each
other.  This allows for a quite explicit treatment of the models.
Levin and Wen studied the properties of physical excitations (which
constitute a complete anyon model, or a unitary \emph{braided} tensor
category in the language of \cite{Kitaev:2005}) by looking at loop
operators commuting with the Hamiltonian.  However, there is no
general representation-theoretic classification of excitations as that
for quantum double models (cf.~section \ref{sec:qdoubles}).

Note in addition that the definition of the plaquette projectors in
Eq.~(\ref{stringnets:plaquettesnfull}) uses $F$-symbols, and in
principle these are defined only for processes with legal vertices,
i.e., satisfying the fusion rules.  One may set them to zero whenever
one of the involved vertices is illegal, however this looks
artificial.  In section \ref{sec:mappingqd} we will see how the
definition of $B^{\text{SN}}_p$ for legal vertices agrees with and is
naturally generalised by that of the quantum doubles for the subclass
of string-nets in the range of our mapping.

\section{\label{sec:mappingqd}%
  Mapping quantum doubles to string nets}

Consider the $\mathrm{D} (G)$ model defined on a planar lattice
$\Lambda$, and perform a basis change at each oriented edge to the
Fourier basis defined by
\begin{equation}\label{qdouble:dtog}
  \ket{ \mu, a, b }
=
  \sqrt{ \frac{ \abs{\mu} }{ \abs{G} } }
  \sum_{g\in G}
  [ D^\mu(g) ]_{ab} \ket{g}
  \; ,
\end{equation}
where $\mu \in \widehat{G}$ runs over the irreducible representations
of $G$, and $D^\mu$ is a fixed matrix realisation of the irreducible
representation $\mu$ (with dimension $\abs{\mu}$.)  Standard
representation-theoretical orthogonality relations imply that
$\widetilde{ \mathcal{B} } = \{ \ket{ \mu, a, b } \}$ is an
orthonormal basis.  The orientation-reversing isomorphism is given by
$\ket{ \mu, a, b }_e \mapsto \ket{ \mu^*, b, a }_{e^*}$.

This change of basis can be interpreted loosely as splitting the local
degrees of freedom into three subspaces, one labelled by the
irreducible representations of $G$, the other two labelled by matrix
elements of these representations (this is not a rigorous
interpretation because the dimensions of the latter subspaces depend
on the irreducible representation; the rigorous statement is the
Peter-Weyl theorem.)  We now argue that the matrix indices are
naturally associated with the beginning and end of an oriented edge,
and that the effect of vertex projectors in equation
(\ref{qdouble:hamiltonian}) is to determine the contraction of these
indices at each vertex, so the degrees of freedom remaining after
imposing vertex projectors are just the irreducible representations of
$G$, for which the model can be interpreted as a string-net model,
with fusion rules stemming from composition of irreducible
representations.

\begin{figure}
\centering
\includegraphics{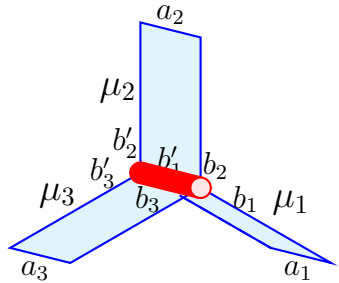}
\caption{\label{fig:wproj}%
  Graphical representation of the projector $A_v^{ \text{QD} }$, or
  more precisely its matrix element $\bra{ \{ \mu_j, \, a_j, \, c_j \}
  } A_v^{ \text{QD} } \ket{ \{ \mu_i, \, a_i, \, b_i \} }$.  Note that
  only the $b$ indices change (into the $c$'s), according to the
  corresponding $3j$ symbol $W_{ \{ c_i \}, \, \{ b_i \} }^{ \{ \mu_i
    \} }$; this is represented by the cylinder.  The trivial
  propagation of the other indices is represented by the rectangular
  faces.}
\end{figure}

Using the inverse change of basis
\begin{equation}\label{qdouble:gtod}
  \ket{g}
=
  \sqrt{ \frac{ \abs{\mu} }{ \abs{G} } }
  \sum_{ \mu \in \widehat{G} }
  \sum_{ a, b }
  [ D^\mu (g) ]_{ab}^*
  \ket{ \mu, a, b }
  \; ,
\end{equation}
it is easy to check that
\begin{equation}\label{qdouble:vprojector_d_full}
  A_v^{ \text{QD} }
  \ket{ \{ \mu_i, a_i, b_i \} }_v
=
  \sum_{ c_1, \dots, c_r}
  W_{ \{ c_i \}, \, \{ b_i \} }^{ \{ \mu_i \} }
  \ket{ \{ \mu_i, a_i, c_i \} }_v
  \; ,
\end{equation}
where
\begin{equation}\label{qdouble:projw}
  W_{ \{ c_i \}, \, \{ b_i \} }^{ \{ \mu_i \} }
=
  \frac{ 1 }{ \abs{G} }
  \sum_{ \ell \in G }
  \prod_i
  [ D^{ \mu_i } ( \ell ) ]_{ c_i b_i }
\end{equation}
is the projector onto the trivial isotypic subspace of $\bigotimes_i
\mu_i$.  In other words, it projects out vertex configurations in
which the irreducible representations $\mu_i$ in the tensor product
$\bigotimes_i \mu_i$ are not coupled to yield the trivial
representation.  This corresponds to the fusion rules in the
string-net model to be identified below.  A graphical interpretation
of $A_v^{ \text{QD} }$ is given in figure \ref{fig:wproj}.
Moreover, since $W^{ \{ \mu_i \} }$ is a projector, it can be split
into a direct sum of orthogonal rank-one projectors, each one of which
corresponds to an inequivalent fusion channel of the $\mu_i$ into the
vacuum:
\begin{equation}\label{qdouble:splitw}
  W_{ \{ a_i \}, \, \{ b_i \} }^{ \{ \mu_i \} }
=
  \sum_A
  W_{ \{ a_i \}, \, \{ b_i \} }^{ \{ \mu_i \}, \, A }
=
  \sum_A
  w_{ \{ a_i \} }^{ \{ \mu_i \}, \, A }
  \bigl( w_{ \{ b_i \} }^{ \{ \mu_i \}, \, A } \bigr)^*
  \; .
\end{equation}
The number of such channels is the trace
\begin{equation}\label{qdouble:tracedelta}
  \Delta_{ \{ \mu_i \} }
=
  \tr W^{ \{ \mu_i \} }
  \; .
\end{equation}
The action of $A_v^{ \text{QD} }$ fixes how the rightmost indices in the ket $\ket{
  \{ \mu_i, \, a_i, \, b_i \} }_v$ should be contracted.  Remember
that we have defined this action assuming that all adjacent edges
point towards $v$. Therefore, these indices correspond naturally to
the ends of the oriented edges.

Now consider the action of the entire set of vertex projectors on the
lattice.  Then all matrix indices are contracted according to the
annihilation channels of the incoming representations.  Hence, if we
consider the ``physical'' Hilbert space to be the surviving subspace
after application of all $A_v^{ \text{QD} }$, the only degrees of freedom left are
precisely the irreducible representations of $G$ living on oriented
edges, with the constraint that representations incident on a given
vertex can fuse to the vacuum~\footnote{%
  Where more than one annihilation channel is available at a vertex,
  another index $A$ as defined in (\ref{qdouble:splitw}) must also be
  specified.  However, for simplicity we will only consider cases
  where all vertex $W$ are at most rank one, $\Delta^{ \{ \mu_i \} }
  \in \{ 0, \, 1 \}$.  These coincide with the `fusion rule deltas' in
  \cite{LevinWen}.}.  We refer to the system in which just irrep
labels are associated with oriented edges, obeying fusion rules, as
the string-net lattice, and we identify a configuration $\ket{ \{ \mu
  \} }_\text{SN}$ there with the state in the original model resulting
from the appropriate contractions of matrix indices with eigenvectors
of the vertex operators as defined in (\ref{qdouble:splitw}):
\begin{equation}\label{qdouble:identifstates}
  \ket{ \{ \mu \} }_\text{SN}
\mapsto
  \sum_{ \{ a, b \} }
  \ket{ \{ \mu, a, b \} }
  \prod_v
  ( w_v )_{ a \ldots b \ldots }
\equiv
  \ket{ \{ \mu \} }_\text{QD}
  \; .
\end{equation}

\begin{figure}
\centering
\includegraphics{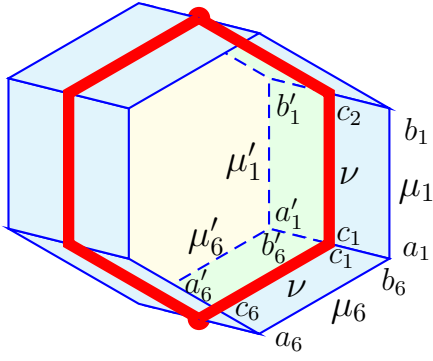}
\caption{\label{fig:bproj}%
  Graphical representation of the projector $B_p^{ \text{QD} } ( \nu
  )$, or more precisely its matrix element $\bra{ \{ \mu'_j, a'_j,
    b'_j \} } B_p^\text{QD} ( \nu ) \ket{ \{ \mu_j, a_j, b_j \} }$ as
  given in eq.~(\ref{qdouble:pprojector_d_full}).  The string-net
  picture of a loop associated with irrep $\nu$ is represented by a
  transversal face (hexagon delimited by bold lines) that interacts
  with the propagation (rectangular faces) of the plaquette via
  $3j$-symbols (bold lines).}
\end{figure}

In order to compute the action of the plaquette projectors
$B_p^\text{QD}$ in the Fourier basis, note that
\begin{equation}\label{qdouble:delta}
  \delta_{g, \, e}
=
  \sum_{ \nu \in \widehat{G} }
  \frac{ \abs{\nu} }{ \abs{G} }
  \chi_\nu (g)
  \; ,
\end{equation}
with $\chi_\nu = \tr D^\nu$ the character of the irreducible
representation $\nu$.  Then
\begin{equation}\label{qdouble:splitbqd}
  B_p^\text{QD}
=
  \sum_{ \nu \in \widehat{G} }
  \frac{ \abs{\nu} }{ \abs{G} }
  B_p^\text{QD} ( \nu )
  \; ,
\end{equation}
with
\begin{align}\label{qdouble:pprojector_d_full}
\nonumber
&  \bra{ \{ \mu'_j, a'_j, b'_j \} } 
  B_p^\text{QD} ( \nu )
  \ket{ \{ \mu_j, a_j, b_j \} } \\
&
\qquad
=
  \prod_{i=1}^s 
  \sqrt{ \abs{ \mu_i } \abs{ \mu'_i } }
  \sum_{ c_1, \dots, c_s }
  \prod_{i=1}^s
  W_{ a_i a'_i c_i, \, b_i b'_i c_{i+1} }^{ \mu_i {\mu'_i}^*\nu }
  \; ,
\end{align}
and the index $i$ understood to be cyclic (see figure
\ref{fig:bproj}).

We assert that this is the correct action of plaquette projectors in
the associated string-net model.  Remember from \cite{LevinWen} and
section \ref{sec:stringnets} that this action is best understood in
the fat lattice picture.  Namely,
\begin{equation}\label{qdouble:bstringnet}
  B_p^\text{SN}
=
  \sum_{ \nu \in \widehat{G} }
  \frac{ \abs{\nu} }{ \abs{G} }
  B_p^\text{SN} ( \nu )
  \; ,
\end{equation}
where we have already identified the quantum dimensions of the labels
as $d_\nu = \abs{\nu}$ and the total quantum dimension as
$\mathcal{D}^2 = \sum_\nu d_\nu^2 = \abs{G}$.  The operator
$B_p^\text{SN} ( \nu )$ is equivalent to creating a loop of label
$\nu$ around the puncture of plaquette $p$ in the fat lattice and then
subsuming it into the original lattice by means of $F$-moves; but in
the case of group representations (and more general representation
theories) $F$-symbols are $6j$-symbols, which can be written entirely
in terms of $W$-projectors (thus, of $3j$ symbols) as defined in
(\ref{qdouble:projw}).  This is explained in appendix~\ref{sec:ident}.

In order to identify the quantum double with a string-net model we
restrict to a trivalent lattice, for definiteness of the honeycomb
type.  We need to show that the action of the $B_p^\text{QD} (\nu)$ on
the reduced QD states $\lvert \cdots \rangle_{\textrm{QD}}$ defined in
equation \ref{qdouble:identifstates} is the same as the action of
$B_p^\text{SN} (\nu)$ on states in the SN lattice.  To this end we
consider a hexagonal plaquette together with its external legs; in the
string-net model its states are labelled by irreducible
representations $\mu_1, \, \ldots, \, \mu_s$ for the edges in a
counterclockwise loop along the plaquette boundary, together with
irrep labels $\alpha_1, \, \ldots, \, \alpha_s$ for the external legs
oriented towards the plaquette.  (The external leg with label
$\alpha_j$ is supposed to end at the vertex where label $\mu_{j-1}$
enters and $\mu_j$ leaves.)  Such a state we denote by $\ket{ \{
  \mu_j, \, \alpha_j \} }_{ \textrm{SN} }$.  The action of the
string-net plaquette operators is \cite{LevinWen}:
\begin{align}\label{qdouble:sandwichbsnnu}
\nonumber
  {}_{ \textrm{SN} }& \bra{ \{ \mu'_j, \, \alpha_j \} }
  B_p^\text{SN} (\nu)
  \ket{ \{ \mu_j, \, \alpha_j \} }_{ \textrm{SN} }
=
  \prod_j
  F^{
     \alpha_j {\mu'_j}^\ast \mu'_{j-1}
  }_{
     \nu \mu_{j-1} \mu_j^\ast
  }
\\
\nonumber
&=
  \prod_j \sqrt{ \abs{\mu'_{j-1}} \abs{\mu_j} }
  \sum_{ a_j, m_j, \bar{m}_j, m'_j, \bar{m}'_j, n_j }
  \bigl(
    w^{ \alpha_j {\mu'_j}^* \mu'_{j-1} }_{ a_j m'_j \bar{m}'_{j-1} }
  \bigr)^*
\\
&
\qquad
 \times
  \bigl(
    w^{
      \nu \mu_{j-1} {\mu'_{j-1}}^*
    }_{
      n_j \bar{m}_{j-1} \bar{m}'_{j-1}
    }
  \bigr)^*
    w^{ \mu_{j-1} \alpha_j \mu_j^* }_{ \bar{m}_{j-1} a_j m_j }
    w^{ {\mu'_j}^* \nu \mu_j }_{ m'_j n_j m_j }
  \; ,
\end{align}
where equation (\ref{qdouble:deff}) from appendix~\ref{sec:ident} has
been used to express $F$ symbols in terms of $3j$ symbols $w$.

In the quantum double model, we define states $\ket{ \{ \mu_j,
  \alpha_j \} }_{ \textrm{QD} }$ for the same system according to rule
(\ref{qdouble:identifstates}); the local Hilbert spaces are full group
algebras labelled by irrep and matrix indices, but the latter are
contracted together with $3j$ symbols $w$.  The action of the quantum
double plaquette operator is
\begin{align}\label{qdouble:sandwichbqdnu}
\nonumber
  {}_{ \textrm{QD} } &\bra{ \{ \mu'_j, \, \alpha_j \} }
  B_p^\text{QD} (\nu)
  \ket{ \{ \mu_j, \alpha_j \} }_{ \textrm{QD} }
\\
\nonumber
&=
  \prod_j
  \sum_{ a_j, m_j, \bar{m}_j }
  \sum_{ a'_j, m'_j, \bar{m}'_j }
  \bigl(
    w^{
      {\mu'_j}^* \mu'_{j-1} \alpha_j
    }_{
      m'_j \bar{m}'_{j-1} a'_j
    }
  \bigr)^*
  w^{ \mu_j^* \mu_{j-1} \alpha_j }_{ m_j \bar{m}_{j-1} a_j }
\\
\nonumber
&
\quad
 \times
  \bra{
    \{ \mu'_j, m'_j, \bar{m}'_j \},
    \{ \alpha_j, b_j, a'_j \}
  }
\\
&
\qquad\qquad
 \times
  B_p^\text{QD} (\nu)
  \ket{
    \{ \mu_j, m_j, \bar{m}_j \},
    \{ \alpha_j, b_j, a_j \}
  }
  \; .
\end{align}

Note that the result is independent of the $b_j$ chosen.  In the right
hand side of the last equation $B_p^\text{QD} (\nu)$ only acts on the
edges of the plaquette according to (\ref{qdouble:pprojector_d_full}),
hence:
\begin{align}\label{qdouble:sandwichbqdnutwo}
\nonumber
  {}_{ \textrm{QD} } & \bra{ \{ \mu'_j, \, \alpha_j \} }
  B_p^\text{QD} (\nu)
  \ket{ \{ \mu_j, \, \alpha_j \} }_{ \textrm{QD} }
\\
\nonumber
&
=
  \prod_j
  \sqrt{ \abs{\mu_j} \abs{\mu'_j} }
  \sum_{ a_j, m_j, \bar{m}_j, m'_j, \bar{m}'_j, c_j }
  \bigl(
    w^{
      {\mu'_j}^* \mu'_{j-1} \alpha_j
    }_{
      m'_j \bar{m}'_{j-1} a_j
    }
  \bigr)^*
\\
&
\quad
 \times
  w^{ \mu_j^* \mu_{j-1} \alpha_j }_{ m_j \bar{m}_{j-1} a_j }
  \,
  W^{
    \mu_j {\mu'_j}^*\nu
  }_{
    m_j m'_j c_j, \bar{m}_j \bar{m}'_j c_{j+1}
  }
  \; ,
\end{align}
which coincides with (\ref{qdouble:sandwichbsnnu}), that is,
\begin{align}\label{conclusion}
\nonumber
  {}_{ \textrm{QD} } \bra{ \{ \mu'_j, \, \alpha_j \} }
& 
  B_p^\text{QD} (\nu)
  \ket{ \{ \mu_j, \, \alpha_j \} }_{ \textrm{QD} }
\\
=
  {}_{ \textrm{SN} }& \bra{ \{ \mu'_j, \, \alpha_j \} }
  B_p^\text{SN} (\nu)
  \ket{ \{ \mu_j, \, \alpha_j \} }_{ \textrm{SN} }
  \; .
\end{align}
Let us comment on the structure of this mapping.  The SN definition of
plaquette operators relies on $F$-symbols, whose extension to
configurations violating vertex conditions is somewhat arbitrary.  By
enlarging the SN local Hilbert spaces introducing matrix degrees of
freedom and going over to the QD Hilbert spaces where edges carry a
full group algebra, we are able to express both plaquette and vertex
operators in a way that recovers the SN definition for the reduced
states defined in equation (\ref{qdouble:identifstates}), but carries
over to the full Hilbert space.  In more concrete terms, we can write
\begin{equation}\label{conclusionforbsn}
  B_p^\text{SN}
\sim
  B_p^\text{QD}
 \otimes
  \bigotimes_{ \text{$v$ around $p$} }
  A_v^\text{QD}
  \; ,
\end{equation}
in the sense that $B_p^\text{SN}$ needs the vertices surrounding the
plaquette to fulfil the fusion rules, and in that space its action can
be identified with that of $B_p^\text{SN}$; incidentally, this
accounts for the fact that the SN plaquette operators are 12-local
while the QD plaquette operators are 6-local.

String-net models obtained from quantum doubles by the Fourier mapping
can be defined naturally for general planar lattices, and not only in
trivalent lattices as a generic SN model.  The reason is that the
vertex projectors have a natural interpretation in group
representation theory, which generalises to $n$-valent vertices: a
vertex configuration is allowed if the tensor product of the incident
irreducible representations contains the trivial representation.

Moreover, group theory also provides us with a natural splitting of
the $F$-symbols according to equation (\ref{qdouble:deff}) in
appendix~\ref{sec:ident}, implying that plaquette projectors act
effectively only on the edges of the plaquettes, since the parts
associated with the external legs have the form of vertex projectors
and act trivially on physical states.

More generally, we have an identification of the superselection
sectors as irreducible representations of (in this case) the
quasi-triangular Hopf algebra $\mathrm{D} (G)$.  Note that the matrix
degrees of freedom $a$, $b$ which must be added to the string-net
lattice to fill the quantum double Hilbert spaces with basis $\{
\lvert \mu a b \rangle \}$ allow us to keep track of the internal
degrees of freedom within the different irreps of $\mathrm{D} (G)$
(e.g., the group element labels for the conjugacy classes defining the
magnetic fluxes, the different vectors for the irreps of the group in
electric charges).

{}From a more abstract point of view, both quantum doubles and their
corresponding string-net models can be seen as a procedure to obtain
an anyon model, that of the physical excitations, which is a unitary
braided tensor category.  This has as objects the superselection
sectors, i.e., the excitations classified by irreducible
representations of $\mathrm{D} (G)$.  This is both obtained starting
with the model defined \`a la QD, i.e., starting with a basis labelled
by group elements (objects of a category $G$) and with the model
defined \`a la SN, with bases labelled by irreps (objects of a
category of representations of $G$).  These categories are equivalent
in the sense that they have the same excitations.

\section{\label{sec:conclusions}%
  Conclusions and outlook}

We have shown explicitly how to identify Kitaev's quantum double
models \cite{Kitaev} with a subclass of the string-net models of Levin
and Wen's \cite{LevinWen}.  The general construction for string nets
can be further simplified in this case due to the interpretation of
the fusion rules in terms of group theory.

As a result, the subclass of SN models corresponding to QD models can
be extended naturally to arbitrary planar lattices; their excitations
can be given a representation-theoretic interpretation at the price of
introducing auxiliary degrees of freedom necessary to keep track of
the internal spaces of the different representations; and the
electric-magnetic duality is recovered, in that plaquette projectors
can be given a natural definition that does not depend on the
completion of $F$-symbols outside the space of recouplings with legal
vertices.  This provides a local characterisation of excitations which
we find satisfactory.

Interestingly, from the point of view of category theory the
construction can be seen as an instance of Morita equivalence, which
stresses the practical importance of these models as laboratories to
provide simple examples of abstract mathematical notions which, in
spite of their importance, are only in their way to become everyday
tools of theoretical physicists.

Let us stress the significance of this construction.  On the one hand,
it is a nontrivial mapping relating the physics of two different
classes of topological models.  We have tried to emphasise the
interplay of physical degrees of freedom which is needed to show this
relationship, and how the smaller local Hilbert space for the
string-net lattice can be naturally enlarged to the local Hilbert
space of the corresponding quantum double model.  On the other hand,
it allows for a clearer picture of the anyons appearing as physical
excitations of the particular class of string nets obtained from our
mapping, and this picture can be extended to more general string-net
models as we mention below.

This construction will help throw light as well on the relations among
different tensor network constructions developed recently to describe
exactly both quantum doubles \cite{VerstraetePower}\cite{AguadoVidal}
and string-net models \cite{Koenig}\cite{Buerschaper}\cite{GuTopo}.
On the other hand, it is necessary to discuss the relation between
Kitaev's ribbon operators \cite{Kitaev} and the loop operators defined
by Levin and Wen \cite{LevinWen}.

The current construction can be extended to models based on local
degrees of freedom where the Hilbert space constitutes a $C^\ast$-Hopf
algebra (as anticipated in \cite{Kitaev}); this generalisation will be
given in \cite{BuerschaperHopf}, where the corresponding relations
among tensor network descriptions will also be discussed.  The case of
models based on weak quasi-Hopf algebras \cite{quasiHopf} is the
subject of work in progress \cite{BuerschaperquasiHopf}.  This will
extend the representation-theoretic approach to excitations to wider
subclasses of string-net models.

\emph{Note added:} After submission of this article, related work
(including a discussion of ribbon operators) was reported in
\cite{Kadar}.

\subsection*{\label{sec:acks}%
  Acknowledgements}

It is a pleasure to acknowledge crucial discussions with Guifr\'e
Vidal and Matthias Christandl.  We thank Ignacio Cirac, Andrew
Doherty, Steve Flammia, Liang Kong, and Zhenghan Wang for relevant
discussions and comments.

\appendix

\section{\label{sec:ident}%
  Expression of $F$-symbols in terms of $3j$ symbols}

It can be shown that the projector onto the trivial representation
subspace of the product $\mu \otimes \nu \otimes \lambda \otimes \rho$
of irreducible representations of group $G$ splits into a sum of
orthogonal projectors associated with the internal channel $\sigma$ in
the coupling scheme $\mu \otimes \nu \stackrel{\sigma}{\rightarrow}
\lambda \otimes \rho$, say, as
\begin{equation}\label{qdouble:fourwintothreew}
  W^{ \mu \nu \lambda \rho }
=
  \bigoplus_{ \sigma \in \widehat{G} }
  \Pi_\sigma^{ \mu \nu, \lambda \rho }
  \; ,
\end{equation}
where the projectors
\begin{equation}\label{qdouble:projontochannel}
  \bigl(
    \Pi_\sigma^{ \mu \nu, \lambda \rho }
  \bigr)_{ m n \ell r, \bar{m} \bar{n} \bar{\ell} \bar{r} }
=
  \abs{\sigma}
  \sum_{ s, \bar{s} }
  W_{ m n s, \bar{m} \bar{n} \bar{s}}^{ \mu \nu \sigma }
  W^{ \lambda \rho \sigma^* }_{ \ell r s, \bar{\ell} \bar{r} \bar{s} }
\end{equation}
are expressed in terms of $W$ connecting three irreducible
representations \footnote{%
  It can be seen easily that $\tr \Pi_\sigma^{ \mu \nu, \lambda \rho }
  = \Delta^{ \mu \nu \sigma } \Delta^{ \lambda \rho \sigma^* }$, so
  these are rank one projectors as long as the three-irrep $W$'s
  are.}. This leads to the definition of the $\hat{F}$ operation as
the change of basis, within the range of $W^{ \mu \nu \lambda \rho }$,
from the states associated with $\Pi_\sigma^{ \mu \nu, \lambda \rho}$
to those of the alternative coupling scheme $\Pi_\tau^{ \rho \mu, \nu
  \lambda }$. Explicitly, the operators read
\begin{equation}\label{qdouble:explicitf}
  {\hat{F}}_{ \lambda \rho \tau }^{ \mu \nu \sigma }
=
  \Pi_\sigma^{ \mu \nu, \lambda \rho }
  \Pi_\tau^{ \rho \mu, \nu \lambda }
\end{equation}
and obviously commute with $W^{ \mu \nu \lambda \rho }$.  From here it
is immediate to check, for instance, that
\begin{equation}\label{qdouble:sumbigf}
  \sum_\tau
  {\hat{F}}_{ \lambda \rho \tau }^{ \mu \nu \sigma }
  {\hat{F}}_{ \nu \lambda \xi }^{ \rho \mu \tau }
=
  \delta_{ \sigma \xi^* }
  \Pi_\sigma^{ \mu \nu, \lambda \rho }
  \; .
\end{equation}

In components, taking into account (\ref{qdouble:splitw}) for rank one
projectors, one has
\begin{equation}\label{qdouble:splitf}
  \bigl(
    {\hat{F}}_{ \lambda \rho \tau }^{ \mu \nu \sigma }
  \bigr)_{ m n \ell r, \, \bar{m} \bar{n} \bar{\ell} \bar{r} }
=
  \bigl(
    p_\sigma^{ \mu \nu, \, \lambda \rho }
  \bigr)_{ r m n \ell} \,
  F_{ \lambda \rho \tau }^{ \mu \nu \sigma } \,
  \bigl(
    p_\tau^{ \rho \mu, \, \nu \lambda }
  \bigr)_{ \bar{m} \bar{n} \bar{\ell} \bar{r}}^\ast
  \; ,
\end{equation}
where
\begin{equation}\label{qdouble:defp}
  \bigl(
    p_\sigma^{ \mu \nu, \, \lambda \rho }
  \bigr)_{ m n \ell r }
=
  \sqrt{ \abs{\sigma} }
  \sum_s
  w_{ m n s }^{ \mu \nu \sigma}
  w_{ \ell r s}^{ \lambda \rho \sigma^\ast }
\end{equation}
are $+1$ eigenvectors of the (rank one) projectors in
(\ref{qdouble:projontochannel}),
\begin{equation}\label{projaspp}
  \bigl(
    \Pi_\sigma^{ \mu \nu, \, \lambda \rho }
  \bigr)_{ m n \ell r, \, \bar{m} \bar{n} \bar{\ell} \bar{r} }
=
  \bigl(
    p_\sigma^{ \mu \nu, \, \lambda \rho }
  \bigr)_{ m n \ell r }
  \bigl(
    p_\sigma^{ \mu \nu, \, \lambda \rho}
  \bigr)_{ \bar{m} \bar{n} \bar{\ell} \bar{r} }^\ast
\end{equation}
and
\begin{equation}\label{qdouble:deff}
  F_{ \lambda \rho \tau }^{ \mu \nu \sigma }
=
  \sqrt{ \abs{\sigma} \abs{\tau} }
  \sum_{ m n \ell r s t }
  \bigl(
    w_{ m n s }^{ \mu \nu \sigma }
  \bigr)^\ast
  \bigl(
    w_{ \ell r s }^{ \lambda \rho \sigma^\ast }
  \bigr)^\ast
  w_{ r m t }^{ \rho \mu \tau }
  w_{ n \ell t }^{ \nu \lambda \tau^\ast }
  \; ,
\end{equation}
for which, for instance,
\begin{equation}\label{qdouble:sumsmallf}
  \sum_\tau
  F_{ \lambda \rho \tau }^{ \mu \nu \sigma }
  F_{ \nu \lambda \xi }^{ \rho \mu \tau }
=
  \Delta_{ \mu \nu \sigma }
  \Delta_{ \lambda \rho \xi }
  \delta_{ \sigma \xi^\ast }
\end{equation}
and
\begin{equation}\label{qdouble:simplef}
  F_{ \lambda \rho \tau }^{ \mu \nu 0 }
=
  \sqrt{ \frac{ \abs{\tau} }{ \abs{\mu} \abs{\lambda} } } \,
  \delta_{ \mu \nu^\ast }
  \delta_{ \lambda \rho^\ast }
  \Delta_{ \mu \lambda^\ast \tau }
\end{equation}
(up to a phase from the square root).

Now the effect of the $\hat{F}$ operators can be interpreted directly
in the string-net lattice by forgetting about the $p$ tensors, whose
role is enforcing physical constraints throughout.  The $F_{ \lambda
  \rho \tau }^{ \mu \nu \sigma }$ are the same as the $F$-symbols in
\cite{LevinWen}.


\end{document}